\documentclass[nofootinbib,twocolumn,showpacs,preprintnumbers,pre,aps]{revtex4-1}

\usepackage{graphicx}
\usepackage{bm}
\usepackage{amsmath}
\usepackage{amssymb}
\usepackage{color}
\usepackage{ulem}

\begin{document}

\title{Dynamics of a bilayer membrane with membrane-solvent \\ 
partial slip boundary conditions}%

\author{Kento Yasuda}

\affiliation{
Department of Chemistry, Graduate School of Science and Engineering,
Tokyo Metropolitan University, Tokyo 192-0397, Japan}

\author{Ryuichi Okamoto}

\affiliation{
Research Institute for Interdisciplinary Science, Okayama University, Okayama 700-8530, Japan}

\author{Shigeyuki Komura}\email{komura@tmu.ac.jp}

\affiliation{
Department of Chemistry, Graduate School of Science and Engineering,
Tokyo Metropolitan University, Tokyo 192-0397, Japan}

\author{Jean-Baptiste Fournier}\email{jean-baptiste.fournier@univ-paris-diderot.fr}

\affiliation{
Universit\'e Paris Diderot, Sorbonne Paris Cit\'e, Laboratoire Mati\`ere et Syst\`emes Complexes (MSC), 
UMR 7057 CNRS, F-75205 Paris, France}

\date{\today}

\begin{abstract}
We discuss the dynamics of a bilayer membrane with partial slip boundary conditions between the monolayers and the bulk fluid.
Using Onsager's variational principle to account for the associated dissipations, we derive the coupled dynamic equations for the membrane height and the excess lipid density.
The newly introduced friction coefficients appear in the renormalized fluid viscosities.
For ordinary lipid bilayer membranes, we find that it is generally justified to ignore the 
effects of permeation and parallel slip at the membrane surface. 
\end{abstract}

\maketitle

Much attention has been paid to artificial lipid bilayer membranes as model systems of biological cell membranes~\cite{AlbertsBook}.
The fluidity of biomembranes is guaranteed mainly by the lipid molecules, which are in the 
liquid crystalline state at physiological temperatures.
Biomembranes exhibit a wide variety of complex phenomena, in both statics and dynamics, 
since lipid densities, membrane deformation and surrounding fluids are coupled to each other.
In early theoretically studies, the relaxation rate of a lipid membrane was discussed by 
regarding it as a tensionless elastic sheet undergoing out-of-plane fluctuations~\cite{Brochard75}.
The membrane relaxation was then shown to be dominated by the bending rigidity and the viscosity 
of the surrounding bulk fluid.

Later on, Merkel
\textit{et al.}~\cite{Merkel89JPhys} and Seifert and Langer~\cite{Seifert93}
considered both the inter-monolayer friction and the two-dimensional (2D) hydrodynamics of each monolayer.
Importantly, they obtained another relaxation mode which is associated with the density 
difference between the two monolayers and is further coupled to the bending mode.
Such a relaxation of the density fluctuation is dominated by the inter-monolayer friction. 
The existence of the predicted compressional mode has been confirmed in the recent 
experiment~\cite{Mell15}. After the work by Seifert and Langer, the bilayer nature of lipid membranes has been explicitly considered for two-component bilayer membranes~\cite{Okamoto16} and spherically closed bilayer vesicles~\cite{Miao02,SachinKrishnan16,SachinKrishnan18}.
More recently, the present authors have discussed the dynamics of a bilayer membrane coupled 
to a 2D cytoskeleton~\cite{Okamoto17}.

In all of these works, a partial slip boundary condition between the two monolayers has been employed~\cite{Seifert93}. 
Conversely, the associated models have always assumed a no-slip boundary condition between the monolayer and the outer bulk fluid (solvent). 
Precisely speaking, this no-slip boundary condition requires that the velocity of the bulk fluid has the following properties at the monolayer surfaces:
(i) the velocity component \textit{normal} to the membrane coincides with the change 
rate of the out-of-plane membrane displacement, so that the bulk fluid cannot permeate 
through the membrane, and 
(ii) the \textit{lateral} velocity component coincides with the fluid velocity of the monolayer.

In this paper, we discuss the dynamics of a bilayer membrane with partial slip boundary conditions 
between the monolayers and the bulk fluid, i.e., we study the case when the above conditions (i) and (ii) are violated.
It is of importance to know which of the slipping modes (inter-monolayer or monolayer-solvent) dominates the membrane dynamics at large-wavenumber excitations. 
We use the framework of Onsager's variational principle~\cite{DoiBook} to obtain the governing hydrodynamic equations~\cite{Fournier15}.
In particular, the friction between the monolayer and the bulk fluid is taken into account 
through newly introduced dissipation functions. 
In order to highlight the effects of the monolayer-solvent partial slip boundary condition, we shall closely follow 
the notations in Ref.~\cite{Fournier15}.

\begin{figure}[tbh]
\begin{center}
\includegraphics[scale=0.40]{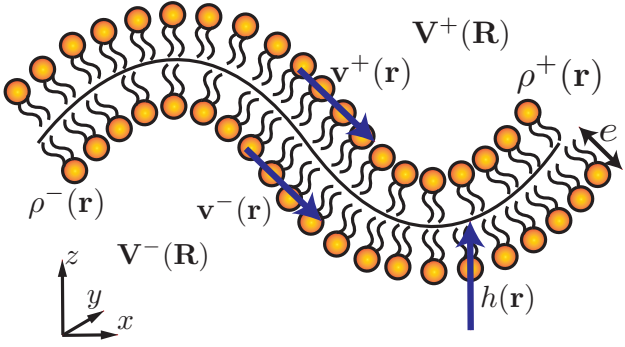}
\end{center}
\caption{
Schematic picture of a laterally compressible bilayer membrane.
The deformation of the membrane is described by the height variable $h(\mathbf r)$
and the excess lipid mass densities of the upper and  the lower monolayers by $\rho^{\pm}(\mathbf r)$.
The lateral velocities of the upper monolayer and the lower monolayer are $\mathbf{v}^{\pm}(\mathbf r,t)$.
The velocities of the bulk fluid in the upper and the lower side of the membrane are 
$\mathbf{V}^{\pm}(\mathbf R,t)$.
}
\label{model}
\end{figure}

As shown in Fig.~\ref{model}, we describe the membrane shape by a height function $h(\mathbf r,t)$, where the 2D vector $\mathbf r=(x,y)$ is a generic point in a reference plane. 
We use $\rho^+(\mathbf r,t)$ and $\rho^-(\mathbf r,t)$ to describe relative excess mass densities of the upper and the lower monolayers, respectively.
The two lipid monolayers are regarded as 2D fluid sheets with a shear viscosity $\eta_2$ and a dilatational viscosity $\lambda_2$.
The lateral velocity of each monolayer is denoted $v_i^\pm(\mathbf r,t)$, with $i\in\{x,y\}$.
The outer bulk fluids are assumed identical on both sides of the membrane and are characterized  by the bulk 
fluid viscosity $\eta$, the pressure 
$P^\pm(\mathbf R,t)$, and the velocity $V_\alpha^\pm(\mathbf R,t)$, where $\alpha\in\{x,y,z\}$ and $\mathbf R=(\mathbf r,z)$ is a three-dimensional vector.

The dynamical equations can be obtained without any ambiguity from the extremalization of the 
Rayleighian~\cite{DoiBook} of the whole system, i.e., the bilayer membrane and the bulk fluid. 
In general, the Rayleighian consists of a dissipation function plus the time derivative of a free energy.
The conservation laws and the boundary conditions are taken into account by introducing 
Lagrange multipliers.
Regarding the bulk solvent as an incompressible fluid, we require $\partial_\alpha V_\alpha^\pm=0$, 
whereas the mass conservation is expressed by $\dot\rho^\pm+\partial_i v_i^\pm=0$.
In this paper, we shall later introduce new friction coefficients between the monolayers and the bulk fluid.

Concerning the boundary condition of the normal component of the velocities, we require that the normal velocities of the upper and  lower monolayers coincide.
However, since we shall use partial slip boundary conditions between the monolayers and the bulk fluid, the velocities of the membrane and the bulk fluid should be different in general. 
Hence we only require the following boundary conditions regarding the bulk fluid velocities:
\begin{align}
V_z^{+}|_{z=0}=V_z^{-}|_{z=0},
\label{eq:BC1}
\end{align}
which ensures that no water is stocked by the membrane.

Next, we consider the dissipation functions of the system.
The dissipation functions corresponding to the bulk fluid $\mathcal P_b^\pm$ and 
the 2D fluid monolayers $\mathcal P_s^\pm$ are given by 
\begin{align}
\mathcal P_b^\pm=\int d^3R\,\eta D_{\alpha\beta}^\pm D_{\alpha\beta}^\pm,
\end{align}
where $D_{\alpha\beta}^\pm=(\partial_\alpha V_\beta^\pm+\partial_\beta V_\alpha^\pm)/2$, and 
\begin{align}
\mathcal P_s^\pm=\int d^2r\,\left(\eta_2 d_{ij}^\pm d_{ij}^\pm+\frac{\lambda_2}{2}d_{ii}^\pm d_{jj}^\pm\right),
\end{align}
where $d_{ij}^\pm=(\partial_i v_j^\pm+\partial_j v_i^\pm)/2$.
Furthermore, the dissipation due to the inter-monolayer friction is given by 
\begin{align}
\mathcal P_i=\int d^2r\,\frac{b}{2}(v_i^+- v_i^-)^2,
\label{disPi}
\end{align}
in which the friction coefficient $b$ between the two monolayers appears.

In contrast to Ref.~\cite{Fournier15}, we shall consider here the dissipation which occurs 
at the boundaries between the monolayers and the bulk fluid.
The friction in the $x,y$-direction is proportional to the relative velocity $v_i^\pm-V_i^\pm|_{z=0}$,
whereas that in the $z$-direction is determined by $\dot h-V_z^+|_{z=0}$, where 
$\dot h$ indicates the time derivative of $h$.
Hence the corresponding dissipation functions can be written as 
\begin{align}
\mathcal P_\parallel^{\pm} & =\int d^2 r\,\frac{b_\parallel}{2}(v_i^\pm-V_i^\pm|_{z=0})^2,\\
\label{Pparallel}
\mathcal P_\perp & =\int d^2 r\,\frac{b_\perp}{2}(\dot h-V_z^+|_{z=0})^2,
\end{align}
where $b_\parallel$ is the friction coefficient in the parallel direction, and $b_\perp$ is that in the normal direction.
These are the new dissipation functions that we consider in this paper.
Notice that the inverse of $b_\perp$ is known as the membrane permeation coefficient~\cite{Manneville01}.

Next we briefly discuss the free energy of a laterally compressible bilayer membrane with finite thickness.
The elasticity of a flexible membrane is generally characterized by a surface tension $\sigma$ and a bending 
rigidity $\kappa$.
Using these quantities, the free energy per unit area of a membrane is given by 
$\frac1 2\sigma(\nabla h)^2+\frac1 2\kappa c^2$ where $c\approx\nabla^2h$ is twice the mean curvature.
We also consider the coupling between the membrane curvature and the lipid density in the 
monolayers~\cite{Seifert93}.
When the membrane has  a positive mean curvature, the upper monolayer is stretched while the lower one is compressed.
The amount of stretching or compression is simply given by $e c$, where $e$ is the distance between the monolayer 
neutral surface and the membrane mid-surface (see Fig.~\ref{model}).
Introducing the monolayer stretching coefficient $k$, one can write down the total free energy as 
\begin{align}
\mathcal H=&\int d^2r\,\left[\frac{\sigma}{2}(\nabla h)^2+\frac{\kappa}{2}(\nabla^2h)^2\right.\nonumber\\&\left.+\frac{k}{2}(\rho^++e\nabla^2h)^2+\frac{k}{2}(\rho^- -e\nabla^2h)^2\right].
\end{align}

The whole set of dynamical equations can be obtained by extremalizing the Rayleighian with respect to all of the 
dynamical variables~\cite{DoiBook}.
The Rayleighian is given by the sum of all the dissipation functions and the time derivative of the  free energy, i.e.,
$\mathcal R=\mathcal P_b^++\mathcal P_b^-+\mathcal P_s^++\mathcal P_s^-
+\mathcal P_i+\mathcal P_\parallel^++\mathcal P_\parallel^-+\mathcal P_\perp+\dot{\mathcal H}$.
The incompressibility condition and the mass conservation condition, as shown before, are taken into account 
by the method of Lagrange multiplier. 
Extremalizing $\mathcal R$ with respect to the fields $v^\pm$, $V^\pm$, $\dot h$, $\dot\rho^\pm$ and the Lagrange multiplier fields, we finally obtain the dynamical equations for a compressible bilayer membrane with partial slip boundary conditions between the monolayers and the bulk fluids.

The equations for the outer bulk fluid are the Stokes equation and the incompressibility condition 
which are given by $-\eta\nabla^2 V_\alpha^\pm+\partial_\alpha P^\pm=0$ and 
$\partial_\alpha V_\alpha^\pm=0$, respectively.
Solving these equations with the use of a 2D Fourier transform of the variables such as 
\begin{align}
X(\mathbf q,z)=\int d^2r \, X(\mathbf R)e^{-i\mathbf q\cdot \mathbf r},
\end{align}
we obtain the bulk fluid velocity as 
\begin{align}
V_z^\pm(\mathbf q,z)& =(A^\pm+B^\pm z)e^{\mp qz},
\label{eq:sol1}\\
V_\parallel^\pm(\mathbf q,z)& =i(B^\pm/q\mp A^\pm\mp B^\pm z)e^{\mp qz},
\label{eq:sol2}
\end{align}
where $V_\parallel^\pm(\mathbf q,z)$ is the component of the bulk fluid velocity parallel 
to the direction of $\mathbf q$, while $A^\pm$ and $B^\pm$ are coefficients yet to be determined.
In the above, we have assumed that $V_\alpha^\pm\to 0$ for $z\to\pm\infty$.
In addition, the pressure is given by $P^\pm(\mathbf q,z)=2\eta B^\pm e^{\mp qz}$.
Notice that the component of the bulk fluid velocity perpendicular to $\mathbf q$, 
denoted by $V_\perp^\pm(\mathbf q,z)$, is completely decoupled from the other components.

The boundary conditions at the membrane, $z=0$, are Eq.~(\ref{eq:BC1}) and 
\begin{align}
&\mp \eta(\partial_z V_i^\pm+\partial_i V_z^\pm)-b_\parallel(v_i^\pm-V_i^\pm)=0, 
\\
&-2\eta(\partial_z V_z^+ -\partial_z V_z^-)+ P^+-P^--b_\perp(\dot h-V_z^+)=0,
\end{align}
which are the force balance equations in the perpendicular and parallel directions to the membrane, 
respectively, in the presence of partial slip effects.
These boundary conditions are now used to obtain the unknown coefficients as 
\begin{align}
A^\pm& =\frac{b_\perp}{b_\perp+4\eta q}\dot h(\mathbf q),
\\
B^\pm& =\pm q\frac{b_\perp}{b_\perp+4\eta q}\dot h(\mathbf q)-iq\frac{b_\parallel}{b_\parallel+2\eta q}v_\parallel^\pm(\mathbf q).
\label{Bpm}
\end{align}
Note that $v_\parallel^\pm(\mathbf q)$ in Eq.~(\ref{Bpm}) is the component of the membrane 
velocity parallel to $\mathbf q$.

The force balance equations for the membrane itself are given by 
\begin{align}
\tilde\kappa\nabla^4h&-\sigma\nabla^2h+ke\nabla^2(\rho^+-\rho^-)+P^+-2\eta\partial_zV_z^+\nonumber\\&-P^-+2\eta\partial_zV_z^-=0,
\end{align}
and 
\begin{align}
-\eta_2\nabla^2v_i^\pm&-(\eta_2+\lambda_2)\partial_i\nabla\cdot v^\pm+b(v_i^\pm-v_i^\mp)\nonumber\\&+k\partial_i(\rho^\pm\pm e\nabla^2h)\mp\eta(\partial_zV_i^\pm+\partial_iV_z^\pm)=0.
\end{align}
Using the solutions for the bulk fluid as calculated above, we finally obtain the dynamical equations for
the membrane height $h$ and the density difference $\rho=\rho^+-\rho^-$ as 
\begin{align}
\left(\begin{matrix}\dot h(\mathbf q,t)\\ \dot \rho(\mathbf q,t)\end{matrix}\right)=-M(q)\left(\begin{matrix} h(\mathbf q,t)\\ \rho(\mathbf q,t)\end{matrix}\right),
\label{eq:Deq}
\end{align}
where
\begin{align}
\renewcommand{\arraystretch}{2.5}
M(q)=\begin{pmatrix}
\displaystyle \frac{\sigma q +\tilde \kappa q^3}{4\eta_\perp(q)} 
& 
\displaystyle -\frac{k e q}{4\eta_\perp(q)} \\ 
\displaystyle -\frac{keq^4}{b+\eta_\parallel(q) q+\eta_sq^2} 
& 
\displaystyle \frac{kq^2}{2[b+\eta_\parallel(q) q+\eta_sq^2]} 
\end{pmatrix}.
\label{Mmatrix}
\end{align}
In the above matrix, we have defined the effective bending rigidity and the effective surface viscosity as
\begin{align}
\tilde \kappa=\kappa+2ke^2,\quad\eta_s=\eta_2+\frac{\lambda_2}{2},
\end{align}
whereas the wavenumber dependent renormalized viscosities are 
\begin{align}
\eta_\perp(q)=\frac{\eta b_\perp}{b_\perp+4\eta q},\quad
\eta_\parallel(q)=\frac{\eta b_\parallel}{b_\parallel+2\eta q}.
\label{effviscosity}
\end{align}
(In Ref.~\cite{Seifert93}, the density difference was defined as $\rho=(\rho^+-\rho^-)/2$.
This leads to a somewhat different appearance of the corresponding relaxation matrix.)
We do not discuss here the dynamics of the density sum $\bar \rho=\rho^++\rho^-$ 
since it is completely decoupled from the other variables~\cite{Seifert93}.
First we note that our result generalizes that of Seifert and Langer~\cite{Seifert93}, because the bulk viscosity $\eta$ is now replaced either by $\eta_\perp(q)$ or $\eta_\parallel(q)$.
In other words, both $\eta_\perp$ and $\eta_\parallel$ reduce to $\eta$ when 
$b_\perp\gg4\eta q$ and $b_\parallel\gg2\eta q$, respectively, corresponding to the 
previous no-slip boundary condition.
Equations (\ref{eq:Deq}) and (\ref{Mmatrix}) are the main result of this paper.

\begin{figure}[tbh]
\begin{center}
\includegraphics[scale=0.35]{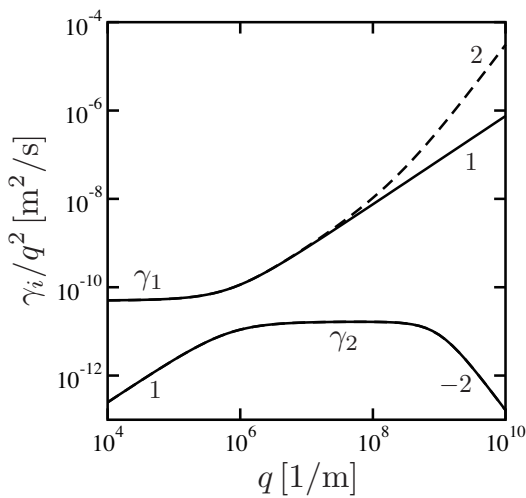}
\end{center}
\caption{
The relaxation rates $\gamma_1$ and $\gamma_2$ as a function of the wavenumber $q$. 
Both are normalized by $q^2$.
The parameter values, given in the text, are those of a tensionless ordinary lipid membrane.
The solid and dashed lines correspond to 
$b_\perp = 10^{12}$~J$\cdot$s/m$^4$ (ordinary membrane) and 
$b_\perp = 10^{6}$~J$\cdot$s/m$^4$ (very permeable membrane), respectively.
The different numbers indicate the slope representing the exponent of the power-law behaviors.
}
\label{figgamma}
\end{figure}

Next we discuss the wavenumber dependence of the relaxation rates, which are the two eigenvalues of the matrix in Eq.~(\ref{Mmatrix}). 
First of all, one can demonstrate that the two eigenvalues are real and positive in all situations.
This property is ensured by the positivity of all the dissipation functions
as well as the positivity of the static parameters such as $\kappa$ and $k$.
In order to see clearly the effects of the partial slip boundary conditions, we consider here the tensionless case and set $\sigma=0$.
The asymptotic analysis of the relaxation matrix yields the following wavenumber dependences
of the two eigenvalues: 
\begin{align}
\gamma_1& \approx
\begin{cases}
(k/2b) q^2, & q\ll q_1, \\  
(\tilde \kappa/4\eta) q^3, & q_1\ll q\ll q_2, \\ 
(\tilde \kappa/b_\perp) q^4, & q_2\ll q,
\end{cases}
\label{gamma1}
\\
\nonumber \\
\gamma_2 & \approx
\begin{cases}
(\kappa/4\eta) q^3, & q\ll q_1, \\  
(k/2b)(\kappa/\tilde \kappa) q^2, & q_1\ll q\ll q_3, \\ 
(k/2\eta_s) (\kappa/\tilde \kappa), & q_3\ll q. 
\end{cases}
\label{gamma2}
\end{align}
Here the three crossover wavenumbers are 
$q_1=2\eta k/(b\tilde \kappa)$,
$q_2=b_\perp/(4\eta)$, and 
$q_3=\sqrt{b/\eta_s}$.
Among these, $q_1$ and $q_3$ were discussed before~\cite{Seifert93}, while $q_2$ introduces 
a new length scale.

The parameters of ordinary lipid membranes have the following order of magnitudes:
$e\approx 10^{-9}$~m, 
$k\approx 0.1$~J/m$^2$, 
$\kappa\approx 10^{-19}$~J,
$b\approx 10^9$~J$\cdot$s/m$^4$,
$\eta_s\approx10^{-9}$~J$\cdot$s/m$^2$, and 
$\eta\approx 10^{-3}$~J$\cdot$s/m$^3$ (pure water)~\cite{Fournier15}.
Note that these values are consistent with the relation $b \approx \eta_s/e^2$.
Likewise, it seems reasonable to estimate the friction coefficient $b_\parallel$ between the monolayer 
and the bulk fluid by using the water viscosity $\eta$ and the water molecular size 
$a \approx 3\times10^{-10}$~m through $b_\parallel \approx \eta/a \approx 10^6$~J$\cdot$s/m$^4$.
As mentioned before, the friction coefficient $b_\perp$ is the inverse of the membrane permeation coefficient; it is known to be $b_\perp \approx 10^{12}$~J$\cdot$s/m$^4$~\cite{Manneville01},
which is far larger than $b_\parallel$.
With these numerical values, the three crossover wavenumbers can be roughly estimated as 
$q_1 \approx 10^6$~m$^{-1}$, 
$q_2 \approx 10^{14}$~m$^{-1}$, and 
$q_3 \approx 10^9$~m$^{-1}$. 
Therefore, we recognize that $q_2$ is beyond the appropriate range for the present theory and that the scaling behavior $\gamma_1 \sim q^4$ for $q \gg q_2$ should not be observable.

\begin{figure}[tbh]
\begin{center}
\includegraphics[scale=0.35]{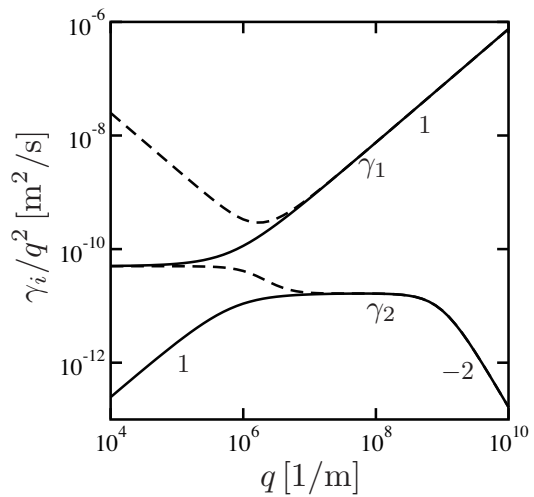}
\end{center}
\caption{
The relaxation rates $\gamma_1$ and $\gamma_2$ as a function of the wavenumber $q$. 
The solid lines show the same data as in Fig.~\ref{figgamma}. 
The dashed line corresponds to the same parameters except for a finite membrane tension 
$\sigma = 10^{-6}$~J/m$^2$. 
The latter affects only the small $q$ behaviors. 
Although the partial slip boundary conditions have been taken into account they also make no visible difference here.
}
\label{figgamma2}
\end{figure}

As for the scale dependent viscosities in Eq.~(\ref{effviscosity}), we expect essentially 
$\eta_\perp \approx \eta$ because $b_\perp$ is very large. 
Besides, $\eta_\parallel$ is generally dependent on $q$ and becomes 
$\eta_\parallel \approx b_\parallel/(2q)$ when $q \gg q^{\ast}$ where 
$q^{\ast} = b_\parallel/(2\eta) \approx 10^8$~m$^{-1}$ for the above parameter values.
Hence, for $q^{\ast} \ll q \ll q_3$, the monolayer friction coefficient is renormalized as 
$b \rightarrow b + b_\parallel$. 
However, since $b_\parallel/b \approx 10^{-3}$ for the above typical parameter 
values, the modification of $b$ due to the partial slip boundary condition may not be observable.
We also note that $q^{\ast}$ is already close to $q_3$.
From these results, one can conclude the effects of partial slip boundary conditions do not show up
in the relaxation dynamics of a compressible bilayer membrane.
In other words, is it justified to neglect both permeation and parallel slip at the membrane surface for ordinary lipid  membranes.

Using the above parameter values, we numerically calculate the two relaxation 
rates $\gamma_1$ and $\gamma_2$ as shown in Fig.~\ref{figgamma} (solid lines).
Although the partial slip boundary conditions are included here, the result is essentially the same as that by Seifert and Langer~\cite{Seifert93} because of the above mentioned reasons.
The wavenumber dependencies of the relaxation rates are in accordance with the asymptotic 
behaviors in Eqs.~(\ref{gamma1}) and (\ref{gamma2}), although we do not see the 
dependence $\gamma_1 \sim q^4$ for $q \gg q_2$ since $q_2$ is too large.
Just for comparison, we have also plotted in Fig.~\ref{figgamma} the relaxation rates for  
$b_\perp \approx 10^{6}$~J$\cdot$s/m$^4$ (dashed line) which corresponds to a very permeable membrane. 
Here $\gamma_1$ increases as $\gamma_1 \sim q^4$ for $q \gg 10^8$~m$^{-1}$. 
In Fig.~\ref{figgamma2} we show the effect of a finite membrane tension of the relaxation rates, 
as presented by the dashed line.

It is instructive to compare here the different dissipation mechanisms introduced 
in the Rayleighian. 
The dissipation due to the inter-monolayer friction is given by Eq.~(\ref{disPi}).
In Fourier space, using the mass conservation law, it can be rewritten as 
\begin{align}
\mathcal P_i  =\int \frac{d^2 q}{(2\pi)^2}\,\frac{b}{2q^2} \vert \dot{\rho}(\mathbf q) \vert^2,
\label{PiFourier}
\end{align}
Similarly, one can Fourier transform the dissipation $\mathcal P_\parallel^{\pm}$ [see  Eq.~(\ref{Pparallel})] due to the friction between the monolayer and the bulk fluid:
\begin{align}
\mathcal P_\parallel^{\pm}  = \int \frac{d^2 q}{(2\pi)^2}\,
\frac{b_\parallel}{q^2} \left( \frac{\eta q}{b_\parallel+2\eta q} \right)^2
\vert \dot{\rho}(\mathbf q) \vert^2.
\label{PparallelFourier}
\end{align}
Looking at the $q$-dependent coefficient of $\vert \dot{\rho}(q) \vert^2$ in the 
integrand, we see that it decays as $q^{-2}$ in Eq.~(\ref{PiFourier}), whereas
in Eq.~(\ref{PparallelFourier}), it is independent of $q$ 
when $q \ll q^{\ast}$, while it also decays as $q^{-2}$ for $q \gg q^{\ast}$.
Hence $\mathcal P_i$ always dominates for $q \ll q^{\ast}$, whereas the sum $b + b_\parallel$ 
contributes to the dissipation for $q \gg q^{\ast}$.
This is consistent with the previous argument on the renormalized friction 
coefficient,  but we note again that $b_\parallel$ is typically much smaller than $b$ for 
ordinary membranes.

In Figs.~\ref{figgamma} and \ref{figgamma2}, we have plotted up to $q=10^{10}$~m$^{-1}$. 
Although directly detecting molecular scale dynamics may not be so easy, we note that a long-wavelength deformation can excite a collection of modes with 
much shorter wavelengths, e.g., when a bilayer membrane is coupled with a cytoskeleton~\cite{Okamoto17}.
This is because the lattice structure of a cytoskeleton breaks lateral continuous translational symmetry and couples Fourier modes with different wave vectors.

Some years ago,  M\"uller and M\"uller-Plathe investigated shear viscosity of a lipid bilayer system 
by using reverse non-equilibrium molecular dynamics simulations~\cite{Muller09}.
They showed that water molecules are less mobile near the lipid headgroups than in the bulk water, 
and the local viscosity of water close to the headgroup interface is several times larger than the bulk 
water viscosity.
This means that the parallel friction coefficient $b_\parallel \approx \eta/a$ can be even larger 
than our estimate (assuming $a$ is the same) because we have used the bulk water viscosity value 
for $\eta$.
In our theory, the increase of $b_\parallel$ leads to the increase of the crossover wavelength $q^{\ast}$.

In conclusion, we have studied the effect on membrane dynamics of partial slip boundary conditions at the 
monolayers-solvent interface. We found that a new regime may appear in the spectrum of the relaxation rates and that the new friction coefficients associated with the partial slip boundary conditions renormalize the solvent viscosity. For ordinary lipid bilayer membranes, however, these effect should not be detectable and it is plainly justified to ignore them. It is nonetheless possible that exotic membranes may someday display the new regimes that we have calculated.

S.K.\ and R.O.\ acknowledge support by Grant-in-Aid for Scientific Research on Innovative 
Areas ``\textit{Fluctuation and Structure}" (Grant No.\ 25103010) from the Ministry of 
Education, Culture, Sports, Science, and Technology (MEXT) of Japan,  and by Grant-in-Aid 
for Scientific Research (C) (Grant No.\ 15K05250) from the Japan Society for the 
Promotion of Science (JSPS).


\end{document}